\documentclass[preprint,prl,aps,showpacs]{revtex4}

\usepackage[dvips]{graphicx}

\begin{document}
%\def\sqr#1#2{{\vcenter{\hrule height.#2pt\hbox{\vrule width.#2pt
%height#1pt \kern#1pt \vrule width.#2pt}\hrule height.#2pt}}}
%\def\square{\mathchoice\sqr64\sqr64\sqr{4.2}3\sqr{3.0}3}
%\draft
\newcommand{\nd}{\noindent}
\newcommand{\be}{\begin{equation}}
\newcommand{\ee}{\end{equation}}
\newcommand{\ben}{\begin{eqnarray}}
\newcommand{\een}{\end{eqnarray}}
\newcommand{\nn}{\nonumber \\}
\newcommand{\ii}{\'{\i}}
\newcommand{\pp}{\prime}
\newcommand{\expq}{e_q}
\newcommand{\lnq}{\ln_q}
\newcommand{\quno}{q-1}
\newcommand{\qunoinv}{\frac{1}{q-1}}
\newcommand{\tr}{{\mathrm{Tr}}}

\draft

%@twocolumnfalse\endcsname
\title{Fisher information and Hamilton's canonical equations}

\author{F.~Pennini}
%\thanks{E-mail:~pennini@fisica.unlp.edu.ar}

\author{A.~Plastino}
%\thanks{E-mail:~plastino@fisica.unlp.edu.ar}

 \address{Instituto de F\'{\i}sica La Plata (IFLP)\\
 Universidad Nacional de La Plata (UNLP) and Argentine National
 Research Council (CONICET)\\ C.C.~727, 1900 La Plata, Argentina}

\begin{abstract}
We show that the mathematical form of the information measure of
Fisher's $I$ for a Gibbs' canonical probability distribution (the
most important one in statistical mechanics) incorporates
important features of the intrinsic structure of classical
mechanics and {\it has a universal form} in terms of ``forces" and
``accelerations", i.e., one that is valid for all Hamiltonian of
the form $T+V$. If the system of differential equations associated
to Hamilton's canonical equations of motion is linear, one can
easily ascertain that the Fisher information per degree of freedom
is proportional to the inverse temperature and to the number of
these degrees. This equipartition of $I$ is also seen to hold in a
simple example involving a non-linear system of differential
equations. 
\end{abstract}

\maketitle

%% 02.50.-r Probability theory, stochastic processes and statistics
%% 05.20.-y Classical statistics mechanics
%% 45.20.-d Formalisms in classical mechanics

\newpage

\section{Introduction}

\nd The last years have witnessed a great deal of activity
revolving around physical applications of Fisher's information
measure (FIM) (as a rather small sample, see for  instance,
  \cite{Frieden,roybook, Renyi,FPPS,Incerteza}).  Frieden and Soffer
  \cite{Frieden} have shown that Fisher's information measure provides one
 with a  powerful variational principle,
  the extreme physical information  one,  that yields most of the canonical
 Lagrangians
   of theoretical physics  \cite{Frieden,roybook}.
  Additionally, $I$ has been shown to provide an interesting characterization
  of
  the ``arrow of time'', alternative to the one associated with Boltzmann's
 entropy
  \cite{pla2,pla4}. Also to be mentioned is a recent
approach to non-equilibrium thermodynamics, based upon Fisher's
measure (a kind of ``Fisher-MaxEnt"), that exhibits definite
advantages over conventional text-book  treatments
\cite{nuestro,nuestro2}.   Thus, unravelling the multiple FIM
facets and their
 links to physics should be of general interest to a vast audience.

  %%%%%%%%%%%%%%%%%%%%%%%%%%%%%%%%%%%%%%%%%%%%%%%%%%%%%%%%%%%%%
  %\subsection{Fisher's Information measure}
  %%%%%%%%%%%%%%%%%%%%%%%%%%%%%%%%%%%%%%%%%%%%%%%%%%%%%%%%%%%%%

  \noindent   R.~A.
  Fisher advanced, already in the twenties, a quite interesting information measure
  (for a detailed study see~\cite{Frieden,roybook}).
  Consider a $\theta\,-\,{\bf z}$ ``scenario" in which we deal
  with a system specified by a physical
  parameter $\theta$,  while ${\bf z}$ is a stochastic variable $({\bf z}\,\in\,\Re^{M})$
  and
  $f_\theta({\bf z})$ the probability density for ${\bf z}$
  (that depends also on  $\theta$).  One makes a
  measurement of
   ${\bf z}$ and
  has to best infer $\theta$ from this  measurement,
   calling the
    resulting estimate $\tilde \theta=\tilde \theta({\bf z})$.
The question is how well $\theta$ can be determined. Estimation
theory~\cite{cramer}
   states that the best possible estimator $\tilde
   \theta({\bf z})$, after a very large number of ${\bf z}$-samples
  is examined, suffers a mean-square error $e^2$ from $\theta$ that
  obeys a relationship involving Fisher's $I$, namely, $Ie^2=1$,
  where the Fisher information measure $I$ is of the form
  \be
  I(\theta)=\int \,\mathrm{d}{\bf z}\,f_\theta({\bf z})\left\{\frac{\partial \ln f_\theta({\bf z})}{
  \partial \theta}\right\}^2  \label{ifisher}.
  \ee
  \noindent  This ``best'' estimator is the so-called {\it efficient} estimator.
  Any other estimator exhibits a larger mean-square error. The only
  caveat to the above result is that all estimators be unbiased,
  i.e., satisfy $ \langle \tilde \theta({\bf z}) \rangle=\,\theta
  \label{unbias}$.   Fisher's information measure has a lower bound:
 no matter what parameter of the system  one chooses to
  measure, $I$ has to be larger or equal than the inverse of the
  mean-square error associated with  the concomitant   experiment.
  This result, $I\,e^2\,\ge \,1,$ is referred to as the
  Cramer--Rao bound \cite{roybook}.
 A particular $I$-case is of great importance:
 that of translation families~\cite{roybook,Renyi},
  i.e., distribution functions (DF) whose {\it
   form} does not change under $\theta$-displacements. These DF
   are shift-invariant (\`a la Mach, no absolute origin for
  $\theta$), and for them
   Fisher's information measure adopts the somewhat simpler appearance
  \cite{roybook}
   \be\label{shift}
  I(shift-invariant)=\int \,\mathrm{d}{\bf z}\,f({\bf z})\,\left\{\frac{\partial \ln
  f({\bf z})}{
  \partial {\bf z}}\right\}^2.
   \ee
\noindent    Fisher's measure is additive~\cite{roybook}. If one
is dealing with phase-space, ${\bf z}$ is a $M=2N$-dimensional
vector, with $N$ ``coordinates" ${\bf r}$ and $N$ ``momenta" ${\bf
p}$. As a consequence, if one wishes to estimate
  phase-space ``location", one confronts an $I({\bf r}+{\bf p})\equiv I({\bf z})$-measure
  and \cite{nuestroPRE}
  \be \label{aditi} I({\bf z})=I({\bf r})+I({\bf p}).\ee
   Here we clearly encounter a  shift-invariance situation.
%%%%%%%%%%%%%%%%%%%%%%%%%%%%%%%%%%%%%%%%%%%%%%%%%%%%%%%%%%%%%%%%
\section{Hamiltonian systems}
%%%%%%%%%%%%%%%%%%%%%%%%%%%%%%%%%

\nd Assume we wish to describe a classical system of $N$ identical
particles of mass $m$ whose  Hamiltonian is of the form  (in
self-explanatory notation) \be \label{TmasV}
\mathcal{H}=\mathcal{T}+\mathcal{V}. \ee A special important
instance is that of
 \be \mathcal{H}({\bf r},{\bf
p})=\sum_{i=1}^N \frac{{\bf p}_i^2}{2 m}+\sum_{i=1}^N V(r_i)
\label{specH}\ee where $({\bf r}_i,{\bf p}_i)$ are the coordinates
and momenta of the  $i$-th particle and $V({\bf r}_i)$ is a
central single-particle potential. Our considerations are {\it not
limited} to Hamiltonians (\ref{specH}), though.

\nd We presuppose that the system is
 in equilibrium at temperature $T$, so that, in the canonical ensemble,
  the probability density reads
\be \rho({\bf r},{\bf p})=\frac{e^{-\beta \mathcal{H}({\bf r},{\bf
p})}}{Z}\label{ro} \ee  with $k$ Boltzmann's constant and ${\bf
r}\,-\,{\bf p}$ denoting two $3N-$dimensional vectors. If $h$
denotes an elementary cell in phase-space, we write, with some
abuse of notation~\cite{pathria1993}, $\mathrm{d} {\bf \tau}
\equiv \mathrm{d}^{3N}{\bf r}\, \mathrm{d}^{3N}{\bf
p}/(N!\,h^{3N})$
 for the pertinent integration-measure while the partition function
 reads~\cite{pathria1993}
 \be
Z=\int\,\frac{\mathrm{d}^{3N}{\bf r}\, \mathrm{d}^{3N}{\bf
p}}{N!\,h^{3N}}\,e^{-\beta \mathcal{H}({\bf r},{\bf p})}. \ee

\nd Remember now Hamilton's celebrated canonical equations \be
\frac{\partial \mathcal{H}({\bf r},{\bf p})}{\partial {\bf
p}}={\bf \dot{r}}; \,\,\,\, \frac{\partial \mathcal{H}({\bf
r},{\bf p})}{\partial {\bf r}}=-{\bf \dot{p}}, \ee so that,
obviously,

  \ben
-kT\, \frac{\partial \ln
  \rho({\bf r},{\bf p})}{\partial {\bf p}}
  &=&{\bf \dot{r}}
 \cr -kT\,\frac{\partial \ln
  \rho({\bf r},{\bf p})}{\partial {\bf r}}
  &=&-{\bf \dot{p}}.
\label{obvious} \een Eqs. (\ref{obvious}) are just the above
referred to canonical equations of Hamilton in a different guise.
We recall them here because the expressions in the left-hand-sides
of (\ref{obvious}) enter the definition of Fisher's measure, if
one expresses $I_\tau$ as a functional of the probability density
$\rho$. Indeed, this FIM has the form (\ref{shift}), so that, in
view of (\ref{aditi}), it acquires the aspect \be I_\tau=\int
\frac{\mathrm{d}^{3N}{\bf r}\, \mathrm{d}^{3N}{\bf p}}{N!\,h^{3N}}
\,\rho({\bf r},{\bf p})\,\mathcal{A}({\bf r},{\bf p}) \ee with \be
\mathcal{A}=a\,\left[\frac{\partial \ln
  \rho({\bf r},{\bf p})}{\partial {\bf p}}\right]^2 +b\,\left[\frac{\partial \ln
  \rho({\bf r},{\bf p})}{\partial {\bf r}}\right]^2
\ee where, for the sake of dimensional balance~\cite{nuestroPRE},
one needs two coefficients $a$ and $b$  that  depend on the
specific  Hamiltonian of the system under scrutiny (see the
following Section). Notice that \be \frac{\partial \ln
  \rho({\bf r},{\bf p})}{\partial {\bf p}}
  =-\beta\,\frac{\partial \mathcal{H}({\bf r},{\bf p})}{\partial {\bf p}}
\ee and \be \frac{\partial \ln
  \rho({\bf r},{\bf p})}{\partial {\bf r}}
  =-\beta \,\frac{\partial \mathcal{H}({\bf r},{\bf p})}{\partial {\bf r}}
\ee so that we can rewrite the Fisher information as \be
(kT)^2I_\tau =a\left\langle \left(\frac{\partial \mathcal{H}({\bf
p},{\bf r})}{\partial {\bf p}}\right)^2 \right\rangle
+b\left\langle \left(\frac{\partial \mathcal{H}({\bf r},{\bf
p})}{\partial {\bf r}}\right)^2\right\rangle. \label{general} \ee
In view of Eqs. (\ref{obvious}), Fisher's measure now becomes \be
I_\tau=\beta^2\,\left\{a\,\left\langle {\bf \dot{r}}^2
\right\rangle +b\,\left\langle {\bf
\dot{p}}^2\right\rangle\right\}.\label{FishHam} \ee  We see that
it incorporates the symplectic structure of classical mechanics.
{\it Eqs. (\ref{general}-\ref{FishHam}) are our main original
results} here: they give the FIM-form for any Hamiltonian of the
form (\ref{TmasV}), a universal ``quadratic" form in the average
values of ``square-accelerations" and ``square-forces". The first
term in the right-hand-side of (\ref{general}) is in fact the same
for all systems for which $\mathcal{T}$ is the conventional
kinetic energy. Note that cyclic canonical coordinates do not
contribute to  $I_\tau$.\vskip 2mm
%%%%%%%%%%%%%%%%%%%%%%%%%%%%%%%%%%%%
\section{Some specially important examples: general quadratic Hamiltonians}

\nd In many important cases \cite{pathria1993} one faces
Hamiltonians of the type

\be \label{linH} \mathcal{H}_L=\sum_{i=1}^N\,\left(  A_i P_i^2+
B_i Q_i^2\right),\ee with $P_i\,-\,Q_i$ generalized momenta and
coordinates, respectively, that lead to  $2N$ differential
equations, the so-called linear Hamiltonian problem~\cite{chaos}.
According to the preceding section we have now

\be \label{Lgral} (kT)^2I_\tau = \sum_{i=1}^N \left[\tilde
A_i\,\left\langle \left(\frac{\partial \mathcal{H}_L}{\partial
P_i}\right)^2 \right\rangle  +\tilde B_i\, \left\langle
\left(\frac{\partial \mathcal{H}_L}{\partial
Q_i}\right)^2\right\rangle\right], \ee it being then obvious that,
in order to achieve a proper dimensional balance, we must have
$\tilde A_i=1/(\beta_0\,A_i)$ and $\tilde B_i=1/(\beta_0\,B_i)$
for all $i$, where $\beta_0$ is a fixed, arbitrary reference
inverse-temperature. Further reflection allows one to see that

\be \label{equi} (kT)^2\,I_\tau= \langle \mathcal{H}_L
\rangle/\beta_0.
 \ee
We know that the equipartition theorem holds for $\langle
\mathcal{H}_L \rangle$~\cite{pathria1993}, so that the mean value
of the energy equals $2NkT$, which entails that it also holds for
the Fisher information \be \label{Fequi} I_\tau=
2N\,(\beta/\beta_0).
 \ee As expected, the information steadily
diminishes as the temperature grows. The divergence at zero
temperature is of no importance, since classical mechanics is not
valid at low enough temperatures.

\vskip 3mm {\bf Free particle} \vskip 3mm
%%%%%%%%%%%%%%%%%%%%%%%%%%%%%%%%%%%
\nd In the case of the $N$ identical free particles, the
Hamiltonian is $\mathcal{H}_i={\bf p}_i^2/2 m$ with
 $i=1\ldots N$. We make here
  the obvious choice $a=m/\beta_0$ (mass). Since
 $\langle {\bf \dot{r}}^2 \rangle=2\langle \mathcal{H} \rangle/m
$ we have \be I_\tau=2  \beta^2\,\left\langle \mathcal{H}/\beta_0
\right\rangle, \ee so that, using the well-known result
(equipartition)~\cite{pathria1993} $\left\langle \mathcal{H}
\right\rangle=3 N /2 \beta$ one obtains \be I_\tau=3 N\,
(\beta/\beta_0). \ee

%%%%%%%%%%%%%%%%%%%%%%%%%%%%%%%%%%%%%%%
\vskip 3mm {\bf $N$ harmonic oscillators} \vskip 4mm
%%%%%%%%%%%%%%%%%%%%%%%%%%%%%%%%%%%%%%%
\nd We consider $N$ identical harmonic oscillators whose
Hamiltonian is $\mathcal{H}_i={\bf p_i}^2/2 m +m \omega {\bf
r_i}^2/2$ with $i=1\ldots N$. It is easy to calculate the
classical mean values in (\ref{FishHam}), which have the form
 $\langle {\bf \dot{r}}^2 \rangle=3N/m\beta$ and
 $\langle {\bf \dot{p}}^2 \rangle=3Nm \omega^2/\beta$.
 In this case the proper choice is $a=m/\beta_0$ and $b=1/m\beta_0\omega^2$,
 so that the information (\ref{FishHam}) adopts the appearance
\be I_\tau=6 N\, (\beta/\beta_0). \ee

\nd In  these two examples we appreciate again the fact that a
Fisher--equipartition emerges, $kT/\beta_0$ per degree of freedom.

\section{Paramagnetic system}

%%%%%%%%%%%%%%%%%%%%%%%%%%%%%%%%%%%%
%%\subsubsection{Simple paramagnetic system}
%%%%%%%%%%%%%%%%%%%%%%%%%%%%%%%%%%%

\nd Consider $N$ magnetic dipoles, each of magnetic moment $\mu$,
in the presence of an external magnetic field ${\bf H}$ of
intensity $H$. These $N$ distinguishable (localized), identical,
mutually non-interacting, and freely orientable dipoles give rise
to a dipole-potential energy ~\cite{pathria1993} \be
\mathcal{H}=-\sum_{i=1}^{N} \,  \overrightarrow{\mu_i} \cdot {\bf
H}=-\mu H\,\sum_{i=1}^{N} \cos \theta_i, \ee where $\theta_i$
gives  the dipole orientation with respect to the field-direction.
Since there is no interaction between our $N$ particles, and both
$I$ and the entropy $S$ are additive quantities, it is permissible
to focus attention on just one generic particle (whose canonical
conjugate variables we call $(\theta, p_{\theta})$) and, at the
end of the calculation, multiply by $N$.
  Hamilton's canonical equations yield then the non-linear equation

  \be \dot{p}_{\theta}=\mu H \sin \theta. \ee
Notice that $p_{\theta}$ does not appear in $\mathcal{H}$, i.e.,
$\partial \mathcal{H}/\partial p_{\theta}=\dot{\theta}=0$, which
entails, of course,  $\langle \dot{\theta} \rangle=0$. Thus,
  in  choosing the  constants entering the definition of $I_\tau$
  (see the preceding Section), we need only to care about $b$,
  with
 $b=1/(\mu\beta_0 H)$.
  The associated  Fisher information is then (Cf. Eq. (\ref{FishHam}))
  \be
I_\tau=\frac{\beta ^2\,N}{\mu H}\, \left\langle \sin^2 \theta
\right\rangle.
  \ee

\nd   We can  easily compute the above mean value using $(sin
\theta \, \mathrm{d} \theta \, \mathrm{d} \varphi)$
  for the elemental solid angle, so that
\be \langle \sin^2 \theta \rangle=\frac{1}{Z}\,\int_0^{2
\pi}\int_0^{\pi}\,e^{\beta \mu H }\,
 \sin^3 \theta\, \mathrm{d}\theta\, \mathrm{d} \varphi \label{int}
\ee where the partition function per particle is of the
form~\cite{pathria1993} \be Z=4\pi\, \frac{\sinh (x)}{x}, \ee with
$x=\mu H/kT$. Evaluating the integral (\ref{int}) explicitly we
obtain \be I_\tau=2\,N\, (\beta/\beta_0)\,L(x), \ee an {\it
original} result, where $L(x)$ is  the well known {\it Langevin
function} \cite{pathria1993} \be \label{lange} L(x)= \coth x
-\frac{1}{x}.\ee Since $L(x)$ vanishes at the origin, so does
$I_\tau$ for infinite temperature, a  result one should expect on
more general grounds~\cite{nuestroPRE}.
 Again, equipartition of Fisher information ensues. Since the
differential equations are not linear in the conjugate (canonical)
variables, the information per degree of freedom does not equal
that of (\ref{Fequi}).

\begin{figure}[!hbp]
\begin{center}
\includegraphics[width=15cm,height=15cm,angle=-90]{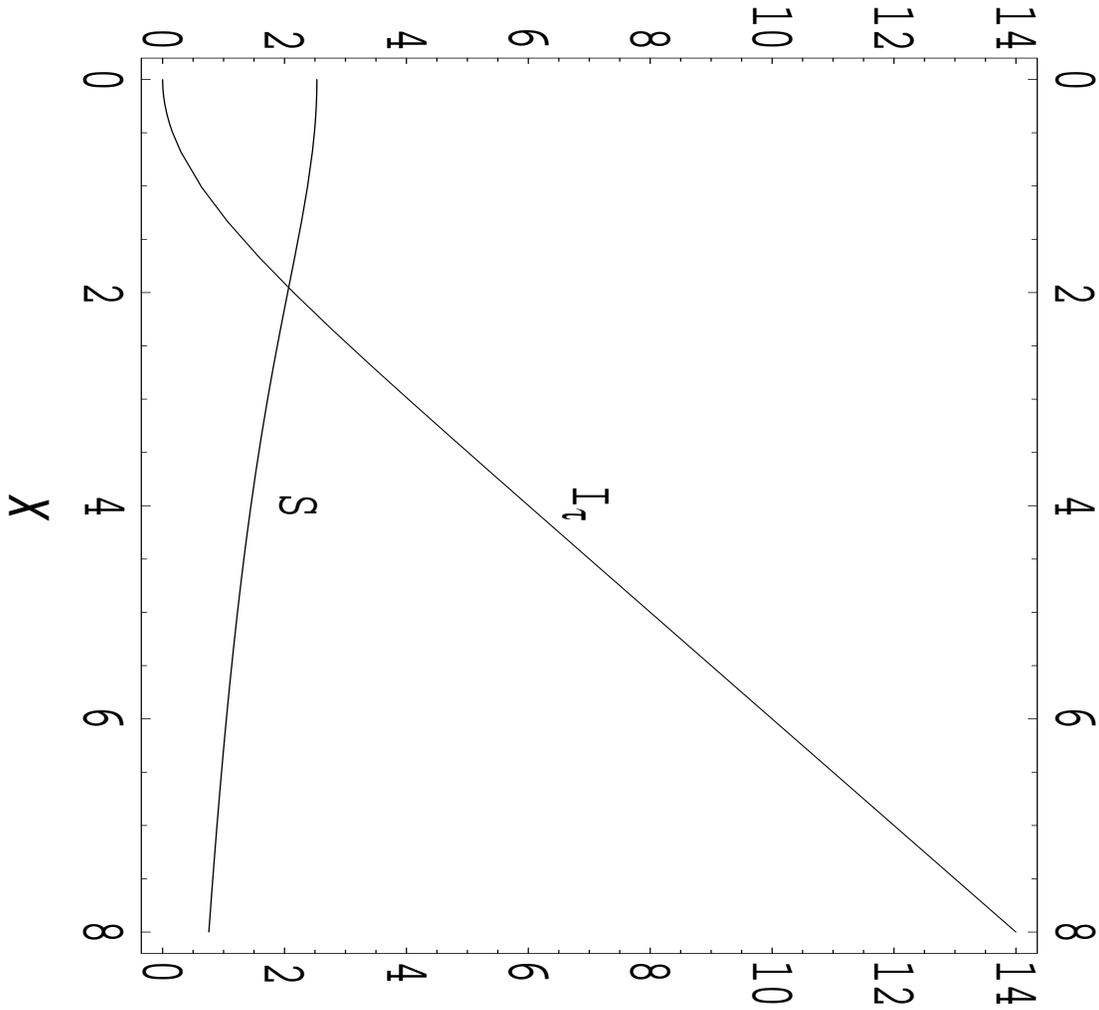}
\caption{We plot i) the entropy $S$ and ii) the Fisher information
$I_\tau$ per particle vs. $x=\mu H/kT$ for $N$ magnetic dipoles,
each of magnetic moment $\mu$, in the presence of an external
magnetic field ${\bf H}$ of intensity $H$. } \label{f.1}
\end{center}
\end{figure}

\section{Conclusions}

We summarize now our results. In this communication we have shown
that
\begin{itemize} \item the mathematical form of the information measure
of Fisher's $I$ for a Gibbs' canonical probability distribution
incorporates important features of the intrinsic structure of
classical mechanics \item it has a universal form in terms of
``forces" and ``accelerations", i.e., one that is valid for any
Hamiltonian of the form $T+V$ \item if the system of differential
equations associated to Hamilton's canonical equations of motion
is linear,
 the amount of Fisher information per degree of freedom is proportional to
the inverse temperature and to the number of these degrees
(equipartition of information!) \item equipartition of $I$ has
been seen to hold also for paramagnetic systems, for which  a
non-linear system of differential equations is involved.
\end{itemize}

\end{document}